\documentclass[pre,superscriptaddress,twocolumn,final,showpacs,showkeys,nobibnotes,titlepage,twoside,10pt]{revtex4-1}
\usepackage{amsmath}
\usepackage{graphicx}
\usepackage{amsfonts}
\usepackage{amssymb}
\usepackage{subfigure}
\usepackage{float}
\usepackage{color}

\begin{document}

\author{M. T. Dang}
\affiliation{Van der Waals-Zeeman Institute, University of Amsterdam, The Netherlands}
\author{V. Chikkadi}
\affiliation{Van der Waals-Zeeman Institute, University of Amsterdam, The Netherlands}
\author{R. Zargar}
\affiliation{Van der Waals-Zeeman Institute, University of Amsterdam, The Netherlands}
\author{D. M. Miedema}
\affiliation{Van der Waals-Zeeman Institute, University of Amsterdam, The Netherlands}
\author{D. Bonn}
\affiliation{Van der Waals-Zeeman Institute, University of Amsterdam, The Netherlands}
\author{A. Zaccone}
\affiliation{Physics Department and Institute for Advanced Study, Technische Universit\"{a}t M\"{u}nchen,
85748 Garching, Germany}
\author{P. Schall}
\affiliation{Van der Waals-Zeeman Institute, University of Amsterdam, The Netherlands}

\begin{abstract}

We develop a free energy framework to describe the response of glasses to applied stress. Unlike crystals, for which the free energy increases quadratically with strain due to affine displacements, for glasses, the nonequilibrium free energy decreases due to complex interplay of non-affine displacements and dissipation. We measure this free energy directly in strained colloidal glasses, and use mean-field theory to relate it to affine and nonaffine displacements. Nonaffine displacements grow with applied shear due to shear-induced loss of structural connectivity. Our mean-field model allows for the first time to disentangle the complex contributions of affine and nonaffine displacements and dissipation in the transient deformation of glasses.
\end{abstract}

\title{Free energy of sheared colloidal glasses}
\maketitle

Unlike crystals that deform reversibly under small applied strain, amorphous materials flow irreversibly even at vanishing applied deformation. This generic relaxation of glasses is due to complex interplay of affine and nonaffine atomic displacements and dissipation, which are direct consequences of the lack of lattice symmetry. While the resulting transient behavior is central to all mechanical properties such as elasticity, time-dependent anelasticity and flow of glasses, its understanding remains a major challenge. An appealing physical picture of relaxation is provided by the free energy landscape of glasses. This concept has remained, however, rather abstract, lacking the direct connection to the atomic displacements. This dilemma is fundamentally connected to the lack of lattice symmetry: in crystals, the lattice point symmetry guarantees local balance of forces by affine displacements, while in glasses, the short-range order precludes force balance by affine displacements alone, leading to additional non-affine motion to restore force balance locally. Disentangling the complex contributions of affine and non-affine displacements and connecting it to the free energy landscape of amorphous materials remains a crucial challenge.

Colloidal glasses allow direct observation of single particle dynamics, and stand out as good model systems to study the microscopic degrees of freedom of amorphous solids. The particles exhibit dynamic arrest due to crowding at volume fractions larger than $\phi _g\sim0.58$, the colloidal glass transition~\cite{Biroli2013}. Recent theoretical~\cite{Aste2004} and experimental work~\cite{Dullens2009,Zargar2013} shows that in these hard-sphere glasses, the particle configurations provide a unique route to the free energy determined by geometry only. This opens up new opportunities to elucidate, experimentally, the relation between microscopic degrees of freedom and important bulk thermodynamic quantities of deformation.

In addition, for the bulk elastic energy, a recent theoretical framework generalizes Born's stability criterion from crystals to amorphous solids. The key ingredient is the lack of a local center of symmetry in amorphous solids, as opposed to crystals: strain-induced forces that do balance by symmetry in centrosymmetric crystals, require additional non-affine displacements in amorphous solids to be relaxed. The resulting framework relates affine and non-affine displacements to the stability of amorphous solids, offering new opportunities to describe their transient deformation.

In this Letter, we provide a generic framework of the transient deformation of glasses by connecting microscopic degrees of freedom to the free energy of deformation. We combine direct measurements of the free energy in sheared hard-sphere glasses with the extended Born's stability framework to link affine and nonaffine displacements to the moduli, stress and free energy of the glass. By applying this formalism to the yielding of colloidal glasses we elucidate how elasticity, vibrational entropy and dissipation interrelate in the deformation. We find that, contrary to the quadratic free energy increase of linear elastic solids, for glasses, the non-equilibrium free energy of deformation decreases with applied strain due to elastic energy being lost in nonaffine displacements and viscous dissipation. We provide a simple particle-scale picture of the transient deformation: the loss of interparticle contacts in the extensional sectors of the shear plane leads to loss of connectivity and proliferation of non-affine displacements, resulting in a marginally stable state when the material eventually yields. This framework allows us to disentangle the complex interplay of rigidity, dissipation, and disorder in the deformation of glasses.

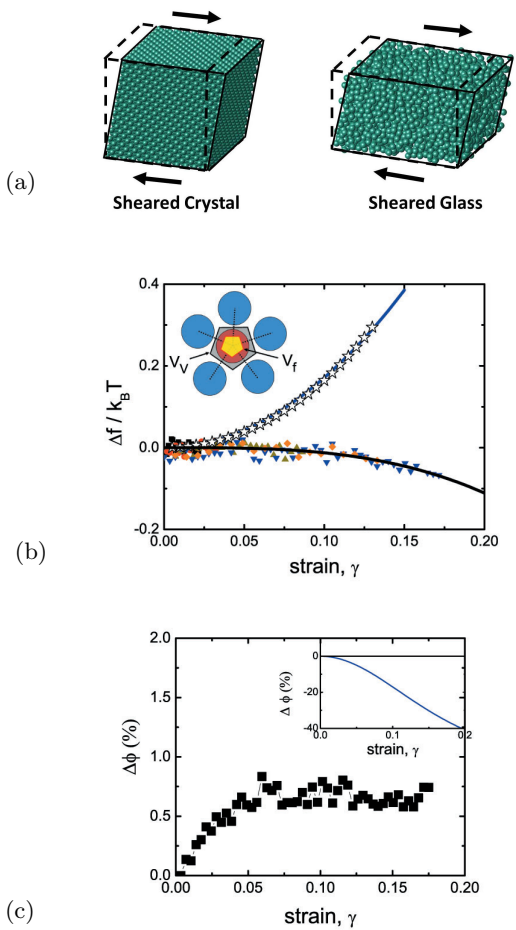
\begin{figure}
\centering
\subfigure
{\includegraphics[width=0.65\columnwidth]{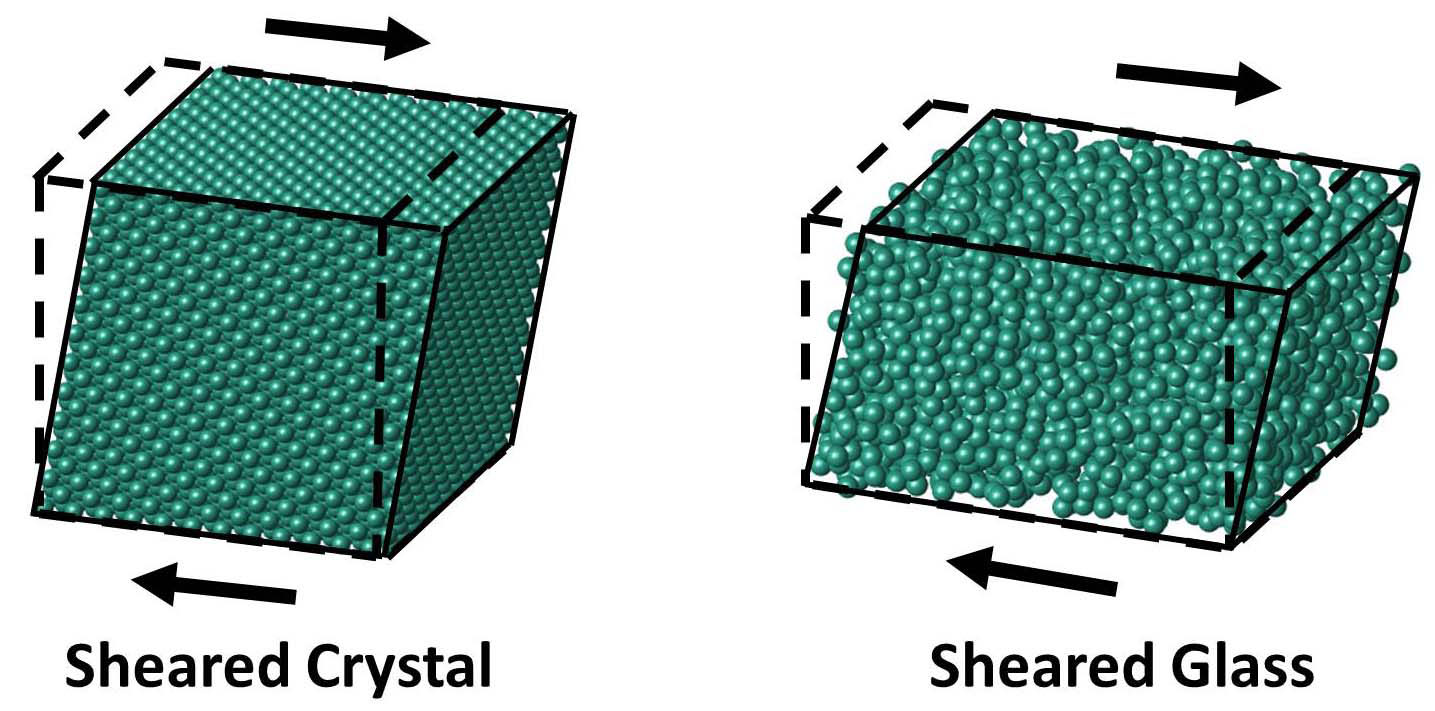}
\label{fig:sketch}}
\begin{picture}(0,0)(0,0)
\put(-200,10){(a)}
\end{picture}
\subfigure
{\includegraphics[width=0.68\columnwidth]{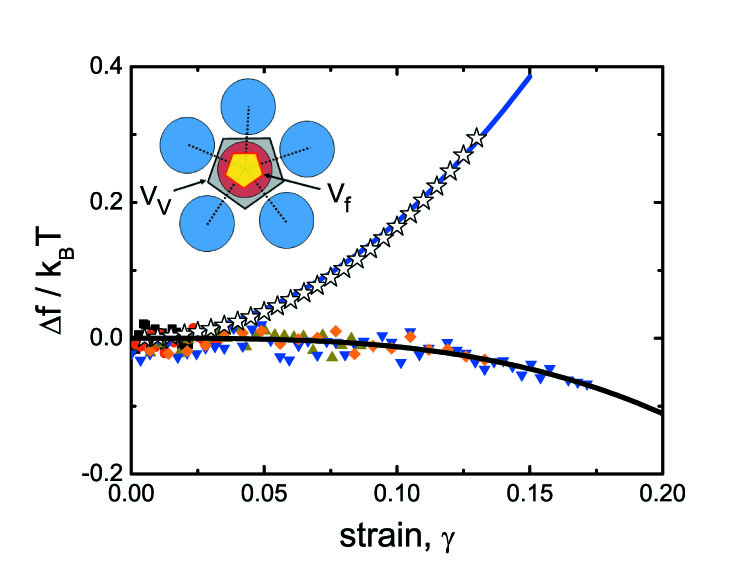}
\label{fig:free_energy}}
\begin{picture}(0,0)(0,0)
\put(-200,10){(b)}
\end{picture}
\subfigure
{\includegraphics[width=0.65\columnwidth]{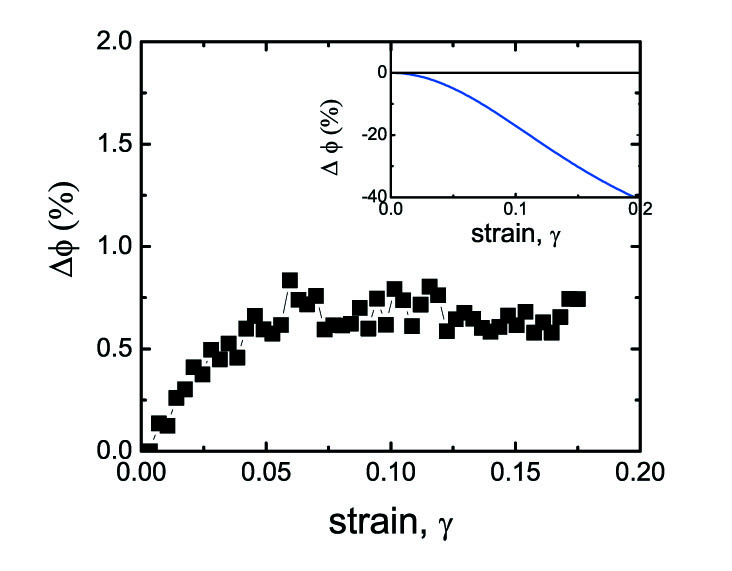}
\label{fig:dilatancy}}
\begin{picture}(0,0)(0,0)
\put(-200,10){(c)}
\end{picture}
\caption{(Color online) Free energy and volume fraction upon deformation of crystal and glass.
\subref{fig:sketch} Sketches of sheared crystal and glasses at zero strain (dash boxes) and strain $\gamma\simeq0.18$ (solid boxes).
\subref{fig:free_energy} Free energy per particle as a function of strain: constructed affine deformation of fcc crystal (stars and up-curved parabolic fit), and experimental deformation of a colloidal glass (filled symbols and down-curved fit according to our model). Symbols indicate shear rates of $\dot{\gamma} = 1.5\times10^{-5} s^{-1}$ (black squares), $3\times10^{-5} s^{-1}$ (red circles), $6\times10^{-5} s^{-1}$ (green triangles), $10^{-4} s^{-1}$ (blue triangles), $2.8\times10^{-4} s^{-1}$ (orange diamonds). Inset: a sketch of Voronoi volume ($v_v$) and free volume ($v_f$) definition.
\subref{fig:dilatancy} Particle volume fraction changes $\Delta\phi$ with strains of the colloidal glass at shear rate $10^{-4} s^{-1}$. Inset: Theoretical volume fraction changes with strains in a simple linear expansion (blue).}
\end{figure}

We use a colloidal glass consisting of sterically stabilized fluorescent polymethylmethacrylate (PMMA) particles with a diameter of $d = 1.5\mu m$, and a polydisperity of $7\%$, suspended in a density and refractive index matching mixture of Cycloheptyl Bromide and Cis-Decalin. A dense glassy suspension with particle volume fraction $\phi \sim 59\%$ is prepared by diluting suspensions centrifuged to a sediment, resulting in a relaxation time of the colloid of $\tau = 2 \times 10^4 s^{-1}$~\cite{Chikkadi2011}. We shear the suspension in between two parallel plates $65\mu m$ apart at constant rates between $\dot{\gamma}=1.5 \times 10^{-5} s^{-1}$ and $\dot{\gamma}=2.8 \times 10^{-4} s^{-1}$, of the order of the inverse relaxation time $\tau^{^-1}$.
Individual particles are imaged using confocal microscopy and their centers are located in three dimensions with an accuracy of $0.03~\mu m$ in the horizontal, and $0.05~\mu m$ in the vertical direction ~\cite{Chikkadi2011}. To measure the free energy~\cite{Zargar2013, Aste2004}, we generate a Voronoi tessellation that divides the space into distinct, non-overlapping convex polyhedra. We then define the free volume $v_{fi}$ of particle $i$ as the volume of a smaller cell generated from the Voronoi cell by moving the faces normally inside over a distance $D = d/2$. The total free energy of the hard-sphere system can be expressed directly in terms of only the free volume as $F\simeq-k_BT\sum_{i=1}^{N} \ln(v_{fi}/\lambda^3)$, where $\lambda=(h^2/2\pi m k_BT)^{1/2}$ is the thermal wavelength. Note that this method is applicable at high densities~\cite{Aste2004} especially for glassy system~\cite{Zargar2013} where the particles are densely packed.

The free energy links the microscopic degrees of freedom directly to the bulk elasticity: For a uniformly sheared crystal, the free energy increases quadratically with applied strain, similar to the elastic deformation of a spring. This is indeed what we find when we construct an affine shear deformation on a computer-generated hard-sphere face-center-cubic (fcc) crystal (see fig.~\ref{fig:sketch}, left), and use our method above to compute the free energy (see open stars and blue fit in Fig.~\ref{fig:free_energy}). Here, the particle volume fraction $\phi=54.5\%$ was preserved using periodic boundary conditions. The curvature of the parabola indicates the elastic modulus, which we determine to be $\mu \sim 34.2 d^3/k_BT$, in good agreement with $\mu \sim 32.4 d^3/k_BT$ determined for $\phi \sim 54.3\%$ in simulations~\cite{FrenkelPronk2003}. In contrast, when we shear a colloidal glass, we measure a surprising decrease of the free energy (filled symbols and black fit in Fig.~\ref{fig:free_energy}). This monotonic decrease indicates that microscopic degrees of freedom other than affine dominate the deformation.

A possible mechanism decreasing the free energy is dilation; glasses can dilate under shear, increasing the free volume, and hence decreasing the free energy.
We thus monitored, by particle counting, the volume fraction upon deformation; we find, however, that the difference $\Delta\phi$ to the quiescent state is small and largely constant in the strain window where the free energy decreases, see Fig.~\ref{fig:dilatancy} (main panel and inset, black line)~\cite{phi}. This is confirmed by independent check of the corresponding pair correlation functions, whose perfect overlap also indicates a closely constant volume fraction, see~\cite{Supmat}. In contrast, by simple linear expansion for marginal elastic solids~\cite{Tighe2014} we estimate a much stronger dilatancy by as much as $\Delta \phi = 40\%$ within the same strain window, see Fig.~\ref{fig:dilatancy} (inset, blue line) and ref.~\cite{Supmat}. This linear expression is defined in terms of a dilatant strain $\epsilon=\frac{1}{2}R_p\gamma^2$ where $R_p=(\frac{dG}{dP})_\gamma - \frac{G}{E} =(\frac{dG}{d\phi}\cdot\frac{d\phi}{dp})_\gamma - \frac{G}{E}$ is the Reynolds coefficient, $E=2G(1+\nu)$ is the Young modulus, $G$ the shear modulus and $\nu=\frac{1}{3}$ the Possion ratio. The resulting strain-dependent volume fraction under constant pressure, $\phi=\phi_0(1+\frac{1}{2}R_p\gamma^2)^{-3}$ with $\phi_0$ the volume fraction of the undeformed state, decreases considerably, in contrast to what is observed in the experiment (Fig.~\ref{fig:dilatancy}).

Hence, the linear expansion grossly overestimates the dilation for all but the smallest deformations and we conclude that the deformation becomes nonlinear very rapidly. Indeed, the distribution of free volumes suggest significant redistribution; such redistributions are caused by non-affine displacements, since affine displacements would leave the overall distribution unaffected. We therefore consider nonaffine displacements to redistribute the microscopic volumes and to lower the free energy of deformation~\cite{Zaccone2011,Zaccone2013}. The main idea is that if the particles are not local centers of symmetry (as is the case in all amorphous solids), there is an imbalance of forces on every particle when a deformation is applied, unlike in crystals with center of inversion symmetry. This additional net force acting on every particle leads to additional nonaffine motions on top of the affine displacements. Because the nonaffine displacements perform internal work against the applied force field, this results in a net negative contribution to the free energy, contrary to the affine displacements that contribute positively.

\begin{figure}
\centering
\mbox{\subfigure[]{
    \includegraphics[width=0.3\columnwidth]{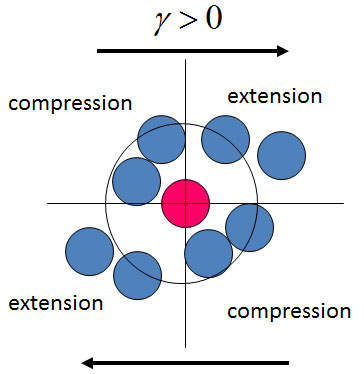}
    \label{fig:model}}    \quad
    \subfigure[]{\includegraphics[width=0.55\columnwidth]{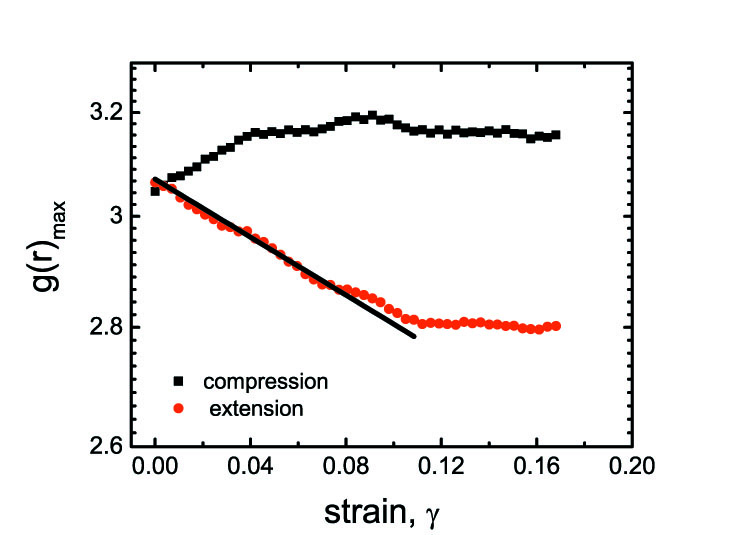}
    \label{fig:gr_max}}}
\caption{(Color online) Shear-induced loss of structural connectivity.
(a) Schematic of nearest neighbor configuration with applied shear. Upon application of shear, the number of particles moving out of the cage in the extension direction is larger than that moving in along the compression direction, leading to net loss of connectivity. (b) Measured maximum of $g(r)$ in the compression (black squares) and extension (red circles) axis (log scale) as a function of strain (linear scale). Black fitting line has slope $-0.9$.}
\end{figure}

To estimate this quantitatively, we first consider the affine part of the free energy, $F_A = \frac{1}{2} G_{A} \gamma^2$, where the shear modulus $G_A=\frac{2}{5\pi}\frac{\kappa\phi}{\sigma}n_{b}$ in the linear regime, according to the Born-Huang theory of lattice dynamics~\cite{Born1954}. Here, $n_b$ is the number of nearest neighbors and $\kappa$ the spring constant associated with a nearest-neighbor bond~\cite{NN}. The number of bonded neighbors is given by the integral of the first peak of $g(r)$, which yields $n_{b}^{0} \approx 12$ for the static hard-sphere glass~\cite{footnote2}, but becomes lower under applied shear, as illustrated in Fig.~\ref{fig:model}: Particles become crowded in the compression sector of the shear plane, whereas they become dilated in the extension sector. Because of strong excluded volume interactions (the nearest neighbors cannot come closer to a selected particle than its excluded volume), the particle increase in the compression sector is comparatively small, and does not completely balance the particle loss in the extension sector, leading to a net loss of particles.

This is indeed what we observe in the experimental pair correlation function resolved along the extension and compression direction (Fig.~\ref{fig:gr_max}): The first maximum $g(r)_{max}$ decreases in the extension direction, indicating particle loss, while in the compression direction, it increases only slightly to saturation. Assuming that the local cage dynamics is governed by the Smoluchowski equation with shear~\cite{Dhont}, we find that the number of nearest neighbors (proportional to the peak of $g(r)$) decreases exponentially with strain, $n_b(\gamma) = n_{b}^{0}\exp(-A\gamma)$ corresponding to a decrease of $g(r)$ in the extension direction $g_c(r) = 3.07\exp(-A\gamma)$ (see \cite{Supmat}), in excellent agreement with the exponential dependence shown in Fig.~\ref{fig:gr_max}. The numerical factor $A$ follows from fit to the experimental data for $g(r)$, and we obtain $A=0.9$. This shear-induced cage-breakdown may be even bigger than what we can here infer on the basis of the static $g(r)$~\cite{NNdecrease}.

The reduced connectivity leads to growing non-affine contributions to the shear modulus. In the spirit of Alexander~\cite{Alexander98}, the nonlinear (non-affine) contribution to the shear modulus can be written as resulting from a Taylor expansion in the free energy up to third order in $\gamma$, $G_{NA}= \frac{2}{5\pi}\frac{\kappa\phi}{\sigma}~(6+C\gamma)$, where $C$ is a phenomenological constant from the non-affine free energy expansion.

\begin{figure}
\centering
{\includegraphics[width=0.65\columnwidth]{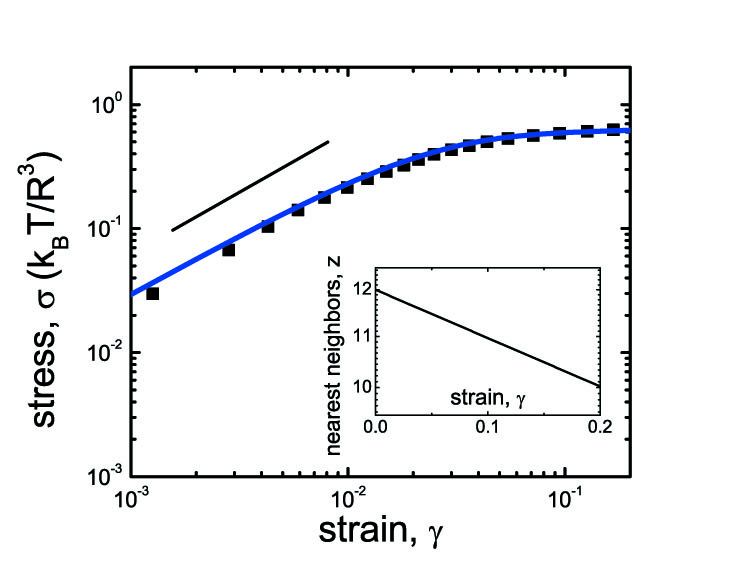}
\caption{(Color online) Comparison of model prediction (blue line) and experimental data (black squares) of stress ($\sigma$) as a function of strain (log-log scale). To measure the stress reliably at these small shear rates, we used smaller particles with diameter $d_1 = 50nm$, resulting in shear stresses larger by a factor $(d/d_1)^3 \sim 3 \times 10^4$. Data was measured at same volume fraction $\phi=0.59$ and similar Peclet number. Black line has slope $1$ to guide the eye. Inset: Theoretical prediction of the average number of nearest neighbors (log scale) as a function of strain (linear scale).}
\label{fig:shear_modulus}}
\end{figure}

We can now evaluate the free energy of deformation quantitatively. The total \emph{nonequilibrium} free energy of deformation is $F (\gamma ) = F_0 + F_{el} (\gamma ) - W_{diss} (\gamma )$, where $F_0$ is the strain-independent, "ground-state" energy of the metastable minimum (inherent structure) of the glass, $W_{diss}(\gamma)$ the energy dissipated due to the microscopic friction and irreversible rearrangements, and $F_{el}=F_A(\gamma)-F_{NA}(\gamma)$ is the elastic (reversible) energy, which can be simply expressed by the shear modulus as $F_{el} = \frac{1}{2}G\gamma^2$, with $G = G_A - G_{NA}$, containing both affine and non-affine contributions. The dissipated energy associated with a continuous ramp of strain at strain rate $\dot{\gamma}$ is $W_{diss}=\int_0^t \sigma(s)\dot{\gamma}ds$, where the stress $\sigma(t)=\dot{\gamma} \int_0^t G(s)ds$~\cite{Zener} with $G$ the (strain dependent) relaxation modulus. Using a standard viscoelastic model, the relaxation modulus has the general form $G(t)=G+G_{R}\exp[-t/\tau]^{\beta}$, where $\tau=\eta/G_{R}$ is the global relaxation time, $\eta$ the viscosity~\cite{Binder} and $G_{R}=G_{0}-G$, with $G_{0}$ the instantaneous (infinite-frequency) shear modulus. For the standard linear viscoelastic solid~\cite{Zener}, one has $\beta=1$, while for many glassy materials the relaxation is stretched exponential with $\beta<1$. Thus, inserting the stretched-exponential expression for $G(t)$ in the integral, and using $t=\gamma/\dot{\gamma}$, one obtains $W_{diss}$ as a function of $\dot{\gamma}$, $\gamma$ and $G_{R}$. Focusing on the limit of very low $\dot{\gamma}$, the leading term in a Taylor expansion around $\dot{\gamma}=0$ is $W_{diss}\approx\frac{1}{2}G_{R}\gamma^{2}$, independent of $\beta$.

\begin{figure}
\centering
{\includegraphics[width=0.65\columnwidth]{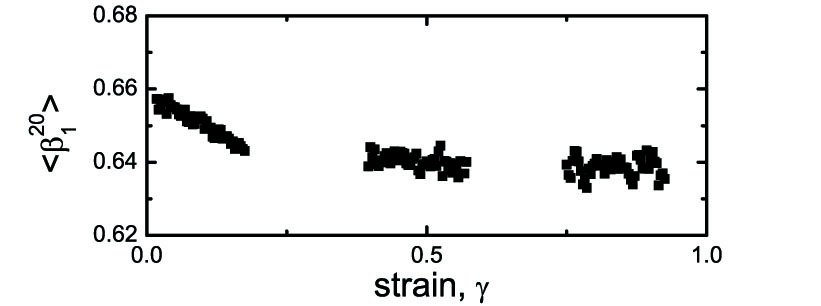}
\includegraphics[width=0.65\columnwidth]{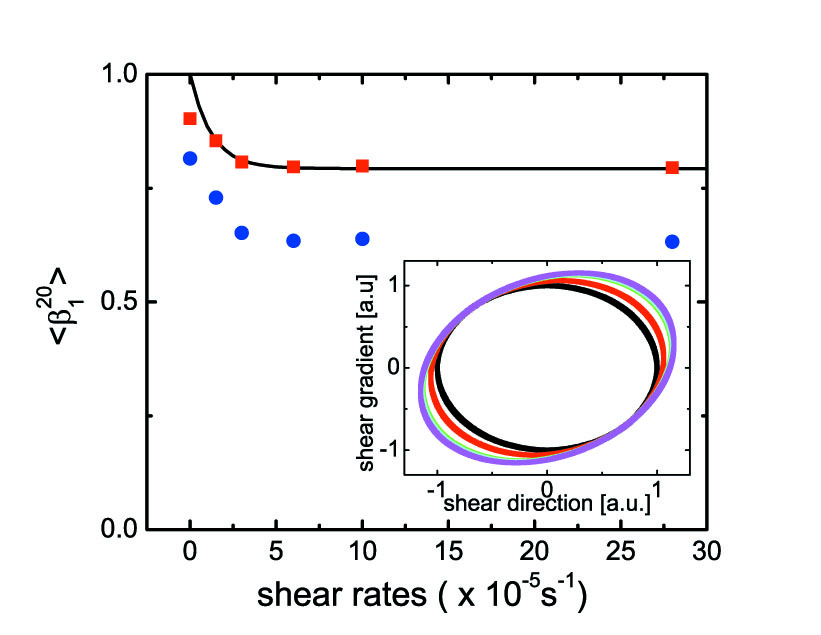}
\put(-200,152){(a)}
\put(-200,0){(b)}
\caption{(Color online) Minkowski measure of cage-distortion underlying the loss of connectivity. Eigenvalue ratio $\beta_1^{20}$ of the Minkowski tensor $W_1^{20}$ indicates elliptical cage distortion. (a) Average ratio $\langle \beta_1^{20} \rangle$ as a function of strain for $\dot{\gamma} = 10^{-4} s^{-1}$. Decreasing values and plateau indicate increasing shear distortion and steady state. (b) Steady state values of $\langle \beta_1^{20} \rangle$ (circles) and $(p/q)$ (squares) as a function of shear rates. Decreasing values indicate increasing elliptical distortion. Inset: Reconstructed effective elliptical cage. Color indicates applied shear rates 0.6 (black), 1.5 (red), 3 (green), 6 (blue), 10 (cyan) and 28 (purple) $\times 10^{-5} s^{-1}$.}
\label{fig:minkowski}}
\end{figure}

Having specified the various contributions, the final expression for the nonequilibrium free energy of the colloidal glass subject to a strain ramp is
\begin{equation}\label{eq:F}
F(\gamma)=F_{0}+\frac{1}{5\pi}\frac{\kappa\phi}{\sigma}[n_{b}^{0}\exp(-A\gamma)-(6+C\gamma)]{\gamma}^2-\frac{1}{2}G_{R}\gamma^{2}
\end{equation}
For a crystal with center-of-inversion symmetry, the non-affine part $F_{NA}$ vanishes and, because typically $G_{A}>G_{R}$, the free energy becomes a monotonic increasing function of $\gamma$, just like the crystal free energy in Fig.~\ref{fig:free_energy} (open stars and blue line).
In amorphous solids, however, non-affine contributions are significant, and ultimately lead to the decreasing behavior of the free energy (jointly with the effect of dissipative terms, also negative). To test our model quantitatively, we simultaneously fit the free energy data in Fig.~\ref{fig:free_energy} using Eq.(1) (black line) and stress-strain curves of a colloidal glass measured independently with a rheometer, see Fig.~\ref{fig:shear_modulus}. We compute the stress as $\sigma =\sigma_{el} + \sigma_{diss}$, where we take $\sigma_{el}=\partial F_{el}/ \partial\gamma$, and $\sigma_{diss}=\eta \dot{\gamma}(1-e^{-\gamma/\dot{\gamma}\tau})$.
We obtain excellent simultaneous fits up to large strains, lending credence to the relation between nonequilibrium free energy and non-affine and dissipative contributions. The predicted number of nearest neighbors (Fig.~\ref{fig:shear_modulus}, inset) demonstrates the loss of connectivity as a function of strain.

We finally highlight the structural distortion of the strained glass
using the Minkowski formalism~\cite{Schroder-Turk2010}.
Unlike $g(r)$ that represents a statistically averaged neighbor distribution, the Minkowski metrics measure directly the local cage anisotropy from higher moments of the Voronoi volume and surface distributions. The tensor $W_1^{20}=\int_S r^2 dA$, where $\bf r$ is the vector from the central particle to its Voronoi neighbors and $S$ is the curvature of the infinitesimal area element $dA$, quantifies elliptical distortions: the average ratio $\langle \beta_1^{20} \rangle$ of the smallest and largest eigenvalues of $W_1^{20}$ is directly related to the ratio of the two semi-axes $p$ and $q$ of an effective elliptical cage according to $\langle \beta_1^{20} \rangle =(p/q)^2$; a value of $\langle \beta_1^{20} \rangle = 1$ indicates an effective isotropic cage, while $\langle \beta_1^{20} \rangle < 1$ indicates elliptical distortion. Indeed, the average of $\beta_1^{20}$ reveals increasing elliptical distortion of the nearest neighbor cage, and saturation at high strains indicating steady state (see fig.~\ref{fig:minkowski} (a)).
The steady-state distortion as a function of strain rates, shown in Fig.~\ref{fig:minkowski} (b), reveals cross-over into shear-rate independent behavior at $\dot{\gamma} \sim \tau^{-1}$. The reconstructed effective elliptical cages (inset) demonstrate the increasing nearest-neighbor distortion. The plateau at $\dot{\gamma} \gtrapprox \tau^{-1}$ is in agreement with our assumption that the cage distortion and number of nearest neighbors become independent of strain rate in the low strain-rate regime investigated here.

In summary, our novel framework connecting microscopic degrees of freedom to the nonequilibrium free energy of glasses allows quantitative description of the transient deformation of glasses far into the nonlinear regime. The sum of the three free energy contributions, affine, non-affine and dissipation, explains the surprising decrease of the nonequilibrium free energy with strain, unlike in crystals. While our hard-sphere colloidal glass allows direct measurement of the free energy and underlying microscopic distortions, the proposed mechanism should apply to molecular glasses as well. For those molecular glasses, the strong hard-core repulsion is replaced by the steep repulsive potential ($\propto r^{12}$) due to Pauli's principle, leading to similar shear-induced loss of connectivity and growth of non-affine dynamics. Hence, interestingly, it is precisely the quantum mechanical principle guaranteeing atomic stability that destabilizes the amorphous material on a larger scale.

\begin{acknowledgments}
This work was supported by the Foundation for Fundamental Research on Matter (FOM) which is subsidized by the Netherlands Organisation for Scientific Research (NWO). We thank Daan Frenkel, Edan Lerner, G. Petekidis, and M. Hoffmann for useful discussions. P. S. acknowledges support by Vidi and Vici fellowships from NWO.
\end{acknowledgments}


\begin{thebibliography}{99}
\bibitem{phi}
Initially, the volume fraction even increases, indicating small compaction rather than dilation. This small amount of compaction can result from the osmotic pressure of the colloidal reservoir that is in direct particle exchange with the sheared colloidal layer.
\bibitem{footnote2} Unlike in the liquid, a good part of the bonded nearest-neighbors in the glass are long-lived, and they contribute the long-time elasticity of the glass~\cite{Leporini}.
\bibitem{NNdecrease}
The static $g(r)$ includes also those neighbours which fluctuate very fast over long distances, hence not contributing to stress-transmission and elasticity. The decrease of neighbours detected from the $g(r)$ is therefore probably an underestimation of the real decrease, which might be stronger if one could separate the contributions of slow (permanent) and fast (non-permanent) neighbours.
\end{thebibliography}
\end{document}